\documentstyle[prl,aps,amsfonts,preprint]{revtex}
 
\def\BF{\leavevmode\null}

\newcommand{\be}{\begin{equation}}
\newcommand{\ee}{\end{equation}}

\newcommand{\bdm}{\begin{displaymath}}
\newcommand{\edm}{\end{displaymath}}
\newcommand{\bea}{\begin{eqnarray}}
\newcommand{\eea}{\end{eqnarray}}
\newcommand{\bean}{\begin{eqnarray*}}
\newcommand{\eean}{\end{eqnarray*}}
\newcommand{\ben}{\begin{enumerate}}
\newcommand{\een}{\end{enumerate}}
\newcommand{\bit}{\begin{itemize}}
\newcommand{\eit}{\end{itemize}}
\newcommand{\la}[1]{\label{#1}}

\newcommand{\eq}[1]{eq.~(\ref{#1})}
\newcommand{\Eq}[1]{Eq.~(\ref{#1})}

\begin{document}
\draft\tighten
\pagestyle{myheadings}
\markright{Review of Speculative ``Disaster Scenarios'' at RHIC}

\setcounter{page}{1}
\title{
Review of Speculative ``Disaster Scenarios'' at RHIC
}

\author{R.L.~Jaffe$^a$, W.~Busza$^{a}$, J.~Sandweiss$^b$,
and F.~Wilczek$^{c}$\\ \null}

\address{a)
Laboratory for Nuclear Science
and Department of Physics\\
Massachusetts Institute of Technology\\
Cambridge, Massachusetts 02139\\[6mm] b) Yale University\\ New
Haven, Connecticut 06520\\[6mm] c) School of Natural Sciences,
Institute for Advanced Study\\ Princeton, New Jersey 08540\\}

\maketitle
\begin{abstract}%
We discuss speculative disaster scenarios inspired by hypothetical new
fundamental processes that might occur in high energy relativistic
heavy ion collisions.  We estimate the parameters relevant to black
hole production; we find that they are absurdly small.  We show that
other accelerator and (especially) cosmic ray environments have
already provided far more auspicious opportunities for transition to a
new vacuum state, so that existing observations provide stringent
bounds.  We discuss in most detail the possibility of producing a
dangerous strangelet.  We argue that four separate requirements are
necessary for this to occur: existence of large stable strangelets,
metastability of intermediate size strangelets, negative charge for
strangelets along the stability line, and production of intermediate
size strangelets in the heavy ion environment.  We discuss both
theoretical and experimental reasons why each of these appears
unlikely; in particular, we know of no plausible suggestion for why
the third or especially the fourth might be true.  Given minimal
physical assumptions the continued existence of the Moon, in the form
we know it, despite billions of years of cosmic ray exposure, provides
powerful empirical evidence against the possibility of dangerous
strangelet production.

\end{abstract}
\bigskip
\thispagestyle{empty}

\narrowtext \newpage

\section{Introduction}

Fears have been expressed that heavy ion collisions at the
Relativistic Heavy Ion Collider (RHIC), which Brookhaven National
Laboratory (BNL) is now commissioning, might initiate a catastrophic
process with profound implications for health and safety.  In this
paper we explore the physical basis for speculative disaster scenarios
at RHIC\cite{JM}.

Concerns have been raised in three general categories: first,
formation of a black hole or gravitational singularity that accretes
ordinary matter; second, initiation of a transition to a lower vacuum
state; and third, formation of a stable ``strangelet'' that accretes
ordinary matter.  We have reviewed the scientific literature,
evaluated recent correspondence, and undertaken additional
calculations where necessary, to evaluate the scientific basis of
these safety concerns.

Our conclusion is that the candidate mechanisms for catastrophe
scenarios at RHIC are firmly excluded by compelling arguments based on
well-established physical laws.  In addition, where the data exists, a
conservative analysis of existing empirical evidence excludes the
possibility of a dangerous event at RHIC at a very high level of
confidence.  Accordingly, we see no reason to delay the commissioning
of RHIC on account of these safety concerns.

 {\BF Considerable attention has been focused on the possibility of placing
a bound on the probability of a dangerous event at RHIC by making a
``worst case'' analysis of certain cosmic ray data\cite{DDH}. We
believe it is reasonable to assume that the laws of physics will not
suddenly break down in bizarre ways when entering a regime that 
actually differs only slightly and in apparently inessential ways
from regimes already well explored.  We will review the work that has
been done on empirical bounds and point out where and how the laws of
physics must be bent in order to avoid very firm bounds on the
probability of a dangerous event at RHIC. No limit is possible if one
allows arbitrarily poor physics assumptions in pursuit of a worst case
scenario.}

Some of the expressed anxiety seems to be based on a misunderstanding
of the nature of high energy collisions: It is necessary to
distinguish carefully between total energy and energy density.  The
total center of mass energy ($E_{\rm CM}$) of gold-gold collisions at
RHIC will exceed that of any existing accelerator.  But $E_{\rm CM}$
is surely not the right measure of the capacity of a collision to
trigger exotic new phenomena.  If it were, a batter striking a major
league fastball would be performing a far more dangerous experiment
than any contemplated at a high energy accelerator.  To be effective
in triggering exotic new phenomena, energy must be concentrated in a
very small volume.

A better measure of effectiveness is the center of mass energy of the
elementary constituents within the colliding objects.  In the case of
nuclei, the elementary constituents are mainly quarks and gluons, with
small admixtures of virtual photons, electrons, and other elementary
particles.  Using the Fermilab Tevatron and the LEP collider at the
European Center for Nuclear Research (CERN), collisions of these
elementary particles with energies exceeding what will occur at RHIC
have already been extensively studied.

What is truly novel {\BF about heavy ion colliders} compared to {\BF other
accelerator environments} is the volume over which high energy
densities can be achieved and the number of quarks involved.  In a
central gold-gold collision, hundreds of quarks collide at high
energies.  Black holes and vacuum instability are generic concerns
that have been raised, and ought to be considered, each time a new
facility opens up a new high energy frontier.  The fact that RHIC
accelerates heavy ions rather than individual hadrons or leptons makes
for somewhat different circumstances.  Nevertheless there are simple,
convincing arguments that neither poses any significant threat.  The
strangelet scenario is special to the heavy ion environment.  It could
have been raised before the commissioning of the AGS or CERN heavy ion
programs.  Indeed, we believe the probability of a dangerous event,
though still immeasureably small, is greater at AGS or CERN energies
than at RHIC. In light of its special role at RHIC, we pay most
attention to the strangelet scenario.

In the remainder of this Introduction we give brief, non-technical
summaries of our principal conclusions regarding the three potential
dangers.  In the body of the paper which follows we consider each
problem in as much detail as seems appropriate.  First, in Section
\ref{cosmics} we present a summary of cosmic ray data necessary to
make empirical estimates regarding vacuum decay and strangelets.
Sections \ref{blackholes}, \ref{falsevacuum}, and \ref{strangelets}
are devoted to gravitational singularities, vacuum decay, and
strangelets respectively.

When we make quantitative estimates of possible dangerous events at
RHIC, we will quote our results as a probability, $\frak{p}$, of a
single dangerous event over the lifetime of RHIC (assumed to encompass
approximately $2\times 10^{11}$ gold-gold collisions over a 10 year
lifetime at full luminosity).  {\BF We do not attempt to decide what is an
acceptible upper limit on $\frak{p}$, nor do we attempt a ``risk
analysis'', weighing the probability of an adverse event against the 
severity of its consequences. } Ultimately, we rely on
compelling physics arguments which, we believe, exclude a
dangerous event beyond any reasonable level of concern\cite{Kent}.

\subsection{Gravitational Singularities}

Exotic gravitational effects may occur at immense densities.
Conservative dimensionless measures of the strength of gravity give
$10^{-22}$ for classical effects and $10^{-34}$ for quantum effects in
the RHIC environment, in units where 1 represents gravitational
effects as strong as the nuclear force.  The theoretical basis for
these estimates is presented in Section \ref{blackholes}.  In fact RHIC
collisions are expected to be less effective at raising the
density of nuclear matter than collisions at lower energies where the
``stopping power'' is greater and existing accelerators have already
probed larger effective energies.  In no case has any phenomenon
suggestive of gravitational clumping, let alone gravitational collapse
or the production of a singularity, been observed.

\subsection{Vacuum Instability}

Physicists have grown quite accustomed to the idea that empty space ---
what we ordinarily call `vacuum' --- is in reality a highly structured
medium, that can exist in various states or phases, roughly analogous
to the liquid or solid phases of water.  This idea plays an important
role in the Standard Model.  Although certainly nothing in our
existing knowledge of the laws of Nature demands it, several
physicists have speculated on the possibility that our contemporary
`vacuum' is only metastable, and that a sufficiently violent
disturbance might trigger its decay into something quite different.  A
transition of this kind would propagate outward from its source
throughout the universe at the speed of light, and would be
catastrophic.

We know that our world is already in the correct (stable) vacuum for
QCD. Our knowledge of fundamental interactions at higher energies, and
in particular of the interactions responsible for electroweak symmetry
breaking, is much less complete.  While theory strongly suggests that
any possibility for triggering vacuum instability requires
substantially larger energy densities than RHIC will provide, it is
difficult to give a compelling, unequivocal bound based on theoretical
considerations alone.

Fortunately in this case we do not have to rely solely on theory;
there is ample empirical evidence based on cosmic ray data.  Cosmic
rays have been colliding throughout the history of the universe, and
if such a transition were possible it would have been triggered long
ago.  Motivated by the RHIC proposal, in 1983 Hut and Rees~\cite{HR}
calculated the total number of collisions of various types that have
occurred in our past light-cone --- whose effects we would have
experienced.  Even though cosmic ray collisions of heavy ions at RHIC
energies are relatively rare, Hut and Rees found approximately
10$^{47}$ comparable collisions have occurred in our past light cone.
Experimenters expect about 2$\times$10$^{11}$ heavy ion collisions in
the lifetime of RHIC. Thus on empirical grounds alone, the probability
of a vacuum transition at RHIC is bounded by 2$\times 10^{-36}$.  We
can rest assured that RHIC will not drive a transition from our vacuum
to another.  We review and update the arguments of Hut and Rees in
Section \ref{falsevacuum} after introducting the necessary cosmic ray
data in Section \ref{cosmics}.

\subsection{Strangelets}

Theorists have speculated that a form of quark matter, known as
``strange matter'' because it contains many strange quarks, might be
more stable than ordinary nuclei.  Hypothetical small lumps of strange
matter, having atomic masses comparable to ordinary nuclei have been
dubbed ``strangelets''.  Strange matter may exist in the cores of
neutron stars, where it is stabilized by intense pressure.

For strange matter to pose a hazard at a heavy ion collider, four
conditions would have to be met:
\begin{itemize}
   \item Strange matter would have to be absolutely stable in bulk at
   zero external pressure.  If strange matter is not stable, it will
   not form spontaneously.

   \item Strangelets would have to be at least metastable for very
   small atomic mass, for only very small strangelets can conceivably
   be created in heavy ion collisions.

   \item It must be possible to produce such a small, metastable
   strangelet in a heavy ion collision.

   \item The stable composition of a strangelet must be {\it
   negatively\/} charged.  Positively charged strangelets pose no
   threat whatsoever.
\end{itemize}
Each of these conditions is considered unlikely by experts in the
field, for the following reasons:
\begin{itemize}

   \item At present, despite vigorous searches, there is no evidence
   whatsoever for stable strange matter anywhere in the Universe.

   \item On rather general grounds, theory suggests that
   strange matter becomes unstable in small lumps due to surface
   effects.  Strangelets small enough to be produced in heavy ion
   collisions are not expected to be stable enough to be
   dangerous.

   \item It is overwhelmingly likely that the most stable
   configuration of strange matter has positive electric charge.

   \item Theory suggests that heavy ion collisions (and hadron-hadron
   collisions in general) are a poor way to produce strangelets.
   Furthermore, it suggests that the production probability is lower
   at RHIC than at lower energy heavy ion facilities like the AGS and
   CERN. Models and data from lower energy heavy ion colliders
   indicate that the probability of producing a strangelet decreases
   very rapidly with the strangelet's atomic mass.

   \item A negatively charged strangelet with a given baryon number is
   much more difficult to produce than a positively charged
   strangelet with the same baryon number because it must contain
   proportionately more strange quarks.

\end{itemize}

To our knowledge, possible catastrophic consequences of strangelet
formation have not been studied in detail before.  Although the
underlying theory (quantum chromodynamics, or QCD) is fully
established, our ability to use it to predict complex phenomena is
imperfect.  A reasonable, conservative attitude is that theoretical
arguments based on QCD can be trusted when they suggest a safety
margin of many orders of magnitude.  The hypothetical chain of events
that might lead to a catastrophe at RHIC requires several independent,
robust theoretical arguments to be wrong simultaneously.  Thus,
theoretical considerations alone would allow us to exclude any safety
problem at RHIC confidently.

However, one need not use theoretical arguments alone.  We have
considered the implications of natural ``experiments'' elsewhere in
the Universe, where cosmic ray induced heavy ion collisions have been
occurring for a long time.  Recent satellite based experiments have
given us very good information about the abundance of heavy elements
in cosmic rays, making it possible to obtain a reliable estimate of
the rate of such collisions.  We know of two domains where empirical
evidence tells us that cosmic ray collisions have not produced
strangelets with disasterous consequences: first, the surface of the
Moon, which has been impacted by cosmic rays for billions of years,
and second, interstellar space, where the products of cosmic ray
collisions are swept up into the clouds from which new stars are
formed.  In each case the effects of a long-lived, dangerous
strangelet would be obvious, so dangerous strangelet production can be
bounded below some limit.  For example, we know for certain that
iron nuclei with energy in excess of $10$ GeV/nucleon (equivalent to
AGS energies) collide with iron nuclei on the surface of the Moon
approximately $6\times 10^{10}$ times per second.  Over the 5 billion
year life of the Moon approximately $10^{28}$ such collisions have
occurred.  None has produced a dangerous strangelet which came to rest
on the lunar surface, for if it had, the Moon would have been
converted to strange matter.  Similarly, we know that the vast number
of heavy ion collisions in interstellar space have not created a
dangerous strangelet that lived long enough to be swept up into a
star\cite{DDH}.  A dangerous strangelet would trigger the conversion
of its host star into strange matter, an event that would resemble a
supernova.  The present rate of supernovae -- a few per millennium per
galaxy -- translate into a strong upper limit on the probability of
long-lived dangerous strangelet production at RHIC.

To translate each of these results into a bound on $\frak{p}$,
it is necessary to model some aspects of strangelet production,
propagation, and decay.  By making sufficiently unlikely assumptions
about the properties of strangelets, it is possible to render both of
these empirical bounds irrelevant to RHIC. The authors of
Ref.~\cite{DDH} construct just such a model in order to discard the
lunar limits: They assume that strangelets are produced only in
gold-gold collisions, only at or above RHIC energies, and only at rest
in the center of mass.  We are skeptical of all these assumptions.  If
they are accepted, however, lunar persistence provides no useful
limit.  Others, in turn, have pointed out that the astrophysical
limits of Ref.~\cite{DDH} can be avoided if the dangerous strangelet
is metastable and decays by baryon emission with a lifetime longer
than $\sim 10^{-7}$ sec.  In this case strangelets produced in the
interstellar medium decay away before they can trigger the death of
stars, but a negatively charged strangelet produced at RHIC could live
long enough to cause catastrophic results.  Under these conditions the
DDH bound evaporates.

We wish to stress once again that we do not consider these
empirical analyses central to the argument for safety at RHIC. The
arguments which are invoked to destroy the empirical bounds from
cosmic rays, if valid, would not make dangerous strangelet production
at RHIC more likely.  Even if the bounds from lunar and astrophysical
arguments are set aside, we believe that basic physics considerations
rule out the possibility of dangerous strangelet production at RHIC.

\section {Heavy Nuclei in Cosmic Rays}
\label{cosmics}

Cosmic ray processes accurately reproduce the conditions planned for
RHIC. Cosmic rays are known to include heavy nuclei and to reach
extremely high energies.  Hut and Rees~\cite{HR} pioneered the use of
cosmic ray data in their study of decay of a false vacuum.  Dar, De
Rujula and Heinz~\cite{DDH} have recently used similar arguments to
study strangelet production in heavy ion collisions.  Here we
summarize data on heavy nuclei (iron and beyond) in cosmic rays and
carry out some simple estimates of particular processes which will
figure in our discussion of strange matter.  In some instances we use
observations directly; elsewhere reasonable extrapolation allows us to
model behavior where no empirical data are available.

We are interested in cosmic ray collisions which simulate RHIC and
lower energy heavy ion facilities like the AGS. Equivalent stationary
target energies range from 10 GeV/nucleon at the AGS to 20 TeV/nucleon
corresponding to the center of mass energy of 100 GeV/nucleon at RHIC.
The flux of cosmic rays has been measured accurately up to total
energies of order $10^{20}$ eV~\cite{Sooth}.  Many measurements of the
abundance of ultraheavy nuclei in cosmic rays at {\bf GeV/nucleon}
energies are summarized in Ref.~\cite{Binns}.  These measurements are
dominated by energies near the lower energy cutoff of 1.5 GeV/nucleon. 
More extensive measurements have been made of the flux of nuclei in
the iron-nickel ($Z=26-28$) group and lighter.  Data on iron are
available up to energies of order 2 {TeV/nucleon}~\cite{Swordy}. 
However, we know of no direct measurements of the flux of nuclei
heavier than the iron-nickel group at energies above 10 {GeV/nucleon}.

Thus data on iron are available over almost the entire energy range we
need.  For nuclei heavier than iron, data are available close to AGS
energies, but not in the {100 GeV/nucleon--20 TeV/nucleon} domain. 
For ultra heavy nuclei at very high energies, we extrapolate existing
data to higher energies using two standard scaling laws, which agree
excellently with available data.
\begin{itemize}
        \item At energies of interest to us, the flux of every species
        which has been measured shows a simple power law spectrum $dF/dE
        \propto E^{-\gamma}$ with $\gamma\approx 2.5-2.7$.  Swordy et
        al.~\cite{Swordy} found this behavior for oxygen, magnesium,
        silicon as well as hydrogen, helium and iron.  The same power law
        is observed at high energies where data are dominated by
        hydrogen.\footnote{At energies above $10^{15}$ eV the power
        $\gamma$ changes abruptly.  This occurs above the energies of
        interest to us.}\cite{Sooth}

        \item At all energies where they have been measured, the relative
        abundance of nuclear species in cosmic rays reflects their
        abundance in our solar system.  [See, for example, Figure 6 in
        Ref.~\cite{Binns}.]  Exceptions to this rule seem to be less than
        an order of magnitude.  If anything, heavy nuclei are
        expected to be relatively more abundant in high energy cosmic
        rays.
\end{itemize}

In light of these facts we adopt the standard idealization that the
$A$ (baryon number or atomic mass) and $E$ (energy per nucleon)
dependence of the flux of primary cosmic rays factors at  {GeV/nucleon
--TeV/nucleon} energies:

\be
    \frac{dF}{dE} = \Gamma(A,E_{0})(E_{0}/E)^{\gamma},
    \la{B.1}
\ee
where $E_{0}$ is some reference energy.    To be conservative we will
usually take $\gamma=2.7$.  The total flux at energies above some
energy $E$ is given by
\be
    F(A,E) = \int_{E}^{\infty}dE' \frac{dF}{dE'} = \frac{E}{\gamma-1}
    \frac{dF}{dE} = \frac{E}{\gamma-1} \Gamma(A,E)
    \la{B.2}
\ee

The units of $dF/dE$ are \{steradians,sec,m$^{2}$, GeV\}$^{-1}$.  The
flux of cosmic rays is very large in these units.  For example, for
iron at {\bf 10 GeV/nucleon}, according to Swordy et al.~\cite{Swordy}
\be
    \frac{dF}{dE}({\rm Fe},10~{\rm GeV}) \equiv \Gamma({\rm Fe}, 10~
    {\rm GeV}) \approx 4\times 10^{-3} \{{\rm ster}\ {\rm sec}\ {\rm
    m}^{2}\ {\rm
    GeV}\}^{-1}.
    \la{B.4}
\ee
Combining all nuclei with $Z>70$ into our definition of ``gold'', we
find an abundance of $\sim 10^{-5}$ relative to
iron.\footnote{Estimates range from $10^{-5}$~\cite{Binns} to as high
as $10^{-4}$~\cite{SSP}.  To be conservative, we choose a value on the
low side.}

We are interested in cosmic ray initiated heavy ion collisions which
have occurred where we can observe their consequences.  Three particular
examples will figure in our subsequent considerations: a) Cosmic ray
collisions with nuclei on the surface of planetoids that lack an
atmosphere, like the Moon;  b) Cosmic ray collisions in
interstellar space resulting in strangelet production at rest with
respect to the galaxy; c) The integrated number of cosmic ray
collisions in our past light cone.

\subsection{Cosmic ray impacts on the moon}

First we consider cosmic rays impinging on the surface of a
planetoid similar to the Moon.  The number of impacts per second with
energy greater than $E$ on the surface of the planet is given by
$8\pi^{2} R^{2} F(A,E)$, where we measure $R$ in units of $R_{\rm moon}$,
\be
    \frac{dN(A,E)}{dt}= 2\times
    10^{14}\frac{\Gamma(A,E)}{\gamma-1}E\left(\frac{R}{R_{\rm
    moon}}\right)^{2}
    \la{B.3}
\ee
For convenience, we use iron with $E=10$ {\bf GeV/nucleon} as our
reference.  From eqs.~(\ref{B.2})--(\ref{B.3}) we find
\be
    \frac{dN(A,E)}{dt} \approx 5\times 10^{12}
    \frac{\Gamma(A, 10\ {\rm GeV})}
    {\Gamma({\rm Fe}, 10\ {\rm GeV})}
    \left(\frac{10\ {\rm GeV }}{E}\right)^{1.7}
    \left(\frac{R}{R_{\rm moon}}\right)^{2}
    \la{B.5}
\ee
This large instantaneous rate makes it possible to obtain useful
limits from cosmic ray collisions with nuclei on the lunar surface.

\subsection{Cosmic ray collisions in space}

Following Ref.~\cite{DDH}, we consider collisions of cosmic rays in
which the center of mass velocity is less than $v_{\rm crit}=0.1$ in
units of $c$.  With this $v_{\rm crit}$ strangelets produced at rest in
the center of mass will have high probability of slowing down without
undergoing nuclear collisions which would destroy them.  The flux
given in \eq{B.1} is associated with a density, $\frac{dn}{dE} =
\frac{4\pi}{c}\frac{dF}{dE}$.  The rate per unit volume for collisions
of cosmic rays with energy per nucleon greater than $E$ in which all
components of the center of mass velocity are less than $v_{\rm crit}$
is given by
\be
    R(E)=2c\sigma f_{\theta}\int_{E}^{\infty}dE_{1}\int_{(1-v_{\rm
    cm})E_{1}}
    ^{(1+v_{\rm crit})E_{1}} dE_{2}\frac{dn}{dE_{1}}\frac{dn}{dE_{2}},
    \la{B.6}
\ee
where $\sigma=0.18 A^{2/3}$ barns is the geometric cross section, and
$f_{\theta} = 4v_{\rm crit}^{2}$ is a geometric factor measuring the
fraction of collisions in which the transverse velocity is less than
$v_{\rm crit}$.  Substituting from \eq{B.1}, and normalizing to
iron-iron collisions at $E=10$ GeV/nucleon, we obtain
\be
    R(E,A) =10^{-45}\left(\frac{10 \
    \hbox{GeV}}{E}\right)^{3.4}\left( \frac{\Gamma(A)}{\Gamma({\rm
    Fe})}\right)^{2}\left(\frac{A}{56}\right)^{2/3} {\rm cm}^{-3}{\rm
    sec}^{-1}
    \la{B.7}
\ee
Although this rate appears very small, these collisions have been
occurring over very large volumes for billions of years.

\subsection{Cosmic ray collisions in our past light cone}

Finally we update the calculation of Hut and Rees of the total number
of high energy collisions of cosmic rays in our past light cone.  The
number of such collisions for cosmic rays with energy greater than $E$
is given by
\begin{equation}
   N\sim 10^{47}\left(\frac{\Gamma({\rm A})}{\Gamma
   ({\rm Fe})}\right)^{2}
   \left(\frac{56}{A}\right)^{2.7}\left(\frac{100~{\rm
   GeV}}{E}\right)^{3.4},
 \label{1.1}
\end{equation}
where we have normalized to iron at $E=100$ {\bf GeV/nucleon}.  The difference
between the extremely small coefficient in \eq{B.7} and the extremely
large coefficient in \eq{1.1} reflects integration over our past
light cone, i.e., over the volume and age of the universe.

\section{Strength of Gravitational Effects}
\label{blackholes}

Two possible sources of novel gravitational effects might in principle
be activated in collisions at RHIC. The first type is connected with
classical gravity, the second type with quantum gravity.

To estimate the quantitative significance of classical gravity, an
appropriate parameter is
\be
   k_{\rm cl} ~\equiv~ {2G M \over R c^2}
   \la{A.1}
\ee
for a spherical concentration of mass $M$ inside a region of linear
dimension $R$, where $G$ is Newton's constant and $c$ is the
speed of light. It is when $k_{\rm cl} \rightarrow 1$ that the escape
velocity from the surface at $R$, calculated in Newtonian gravity,
becomes equal to the speed of light.  The same parameter,
$2GM/c^{2}$, appears in the general relativistic line element
\begin{equation}
    ds^2 ~=~
    c^2 dt^2 (1 - {2GM\over rc^2})
    - {dr^2\over 1 - {2GM\over rc^2}} - r^2  d^{\,2}  \Omega
    \la{A.2}
\end{equation}
outside a spherical concentration of mass $M$.  In this language, it
is when $k_{\rm cl}= 1$ that a horizon appears at $R$, and the body is
described as a black hole.

Now for RHIC we obtain a very conservative upper bound on $k_{\rm cl}$
by supposing that all the initial energy of the collision becomes
concentrated in a region characterized by the Lorentz-contracted
nuclei with a Lorentz contraction factor of $10^{-2}$.  We are being
extremely conservative by choosing the largest possible mass and the
smallest possible distance scale defined by the collision, and also by
ignoring the effect of the electric charge and the momentum of the
constituents, which will resist any tendency to gravitational
collapse.  Thus our result will provide a bound upon, not an estimate
of, the parameters that might be required to have a realistic shot at
producing black holes.

With $M = 10^4$ Gev/$c^2$ and $R = 10^{-2}\times 10^{-13}$ cm, we
arrive at $k_{\rm cl} = 10^{-22}$.  The outlandishly small value of
this over-generous estimate makes it pointless to attempt refinements.

To estimate the quantitative significance of quantum gravity, we
consider the probability to emit the quantum of gravity, a graviton.
It is governed by
\be
   k_{\rm qu}~ \equiv~ {G E^2  \over \hbar c^5}~,
   \la{A.3}
\ee
where $\hbar$ is Planck's constant and $E$ is the total
center-of-mass  energy of collision.  For collisions between
elementary particles at RHIC, we should put $E\approx 200$ GeV.  This
yields $k_{\rm qu} \approx 10^{-34}$.  Once again, the tiny value of
$k_{\rm qu}$ makes it pointless to attempt refinements of this rough
estimate.  Of course higher-energy accelerators than RHIC achieve
larger values of $k_{\rm qu}$, but for the foreseeable future values
even remotely approaching unity are a pipe dream.

\section {Decay of the False Vacuum}
\label{falsevacuum}

Hut and Rees first examined the question of vacuum stability in
1983\cite{HR}. They reasoned that the transition to the true vacuum,
once initiated, would propagate outward at the speed of light.  Thus
our existence is evidence that no such transition occured in our past
light cone.  Hut and Rees then estimated the total number of cosmic ray
collisions in the RHIC energy regime which have occured in our past
light cone.  They used data on cosmic ray fluxes that have
subsequently been confirmed and updated.  Not knowing which would be
more effective at triggering a transition, Hut and Rees looked both at
proton-proton collisions and collisions of heavy nuclei.  Cosmic ray
data on proton fluxes go up to energies of order $10^{\,20}$
eV~\cite{Sooth}.  They conclude that proton-proton collisions with a
center of mass energy exceeding $10^{\,8}$ TeV have occurred so
frequently in our past light cone that even such astonishingly high
energy collisions can be considered safe.

For heavy ions, Hut and Rees derived an estimate of the number of
cosmic ray collisions in our past light cone.  We have updated their
result in \eq{B.7}, and normalized it so that the coefficient
$10^{47}$ equals the number of iron-iron collisions at a center of
mass energy exceeding 100 GeV/nucleon.  The abundance of iron in
cosmic rays has now been measured up to energies of order 2
TeV/nucleon~\cite{Swordy} and agrees with the estimate used by Hut and
Rees.  This result translates into a bound of $2\times 10^{-36}$ on,
$\frak{p}$, the probability that (in this case) an iron-iron collision
at RHIC energies would trigger a transition to a different vacuum
state.  While we do not have direct measurements of the fractional
abundance of elements heavier than iron in cosmic rays of energy of
order 100 GeV/nucleon, we do have good measurements at lower energies,
where they track quite well with the abundances measured on earth and
in the solar system.  For ``gold'' (defined as $Z>70$) at lower
energies $\Gamma({\rm Au})/\Gamma({\rm Fe})\approx 10^{-5}$, leading
to a bound, ${\frak p}<2\times 10^{-26}$ on the probability that a
gold-gold collision at RHIC would lead to a vacuum transition.  Even
if this estimate were off by many orders of magnitude, we would still
rest assured that RHIC will not drive a transition from our vacuum to
another.

Since the situation has not changed significantly since the work of
Hut and Rees, we do not treat this scenario in more detail here.  The
interested reader should consult Hut's 1984 paper for further details
\cite{HR}.

\section{Strangelets and Strange Matter}
\label{strangelets}

The scientific issues surrounding the possible creation of a
negatively charged, stable strangelet are complicated.  Also, it
appears that if such an object did exist and could be produced at
RHIC, it might indeed be dangerous.  {\BF Therefore we wish to give this
scenario careful consideration.}

This section is organized as follows.  First we give a pedagogical
introduction to the properties of strangelets and strange matter.
Second we discuss the mechanisms that have been proposed for producing
a strangelet in heavy ion collisions.  We examine these mechanisms and
conclude that strangelet production at RHIC is extremely unlikely.
Nevertheless, we go on to discuss what might occur if a stable,
negatively charged strangelet could be produced at RHIC. In light of
the possible consequences of production of a stable negatively charged
strangelet, we shall refer to such an object as a ``dangerous''
strangelet.

We then turn to the cosmic ray data.  We obtain strong bounds on the
dangerous strangelet production probability at RHIC from physically
reasonable assumptions.  We also describe the ways in which these
bounds can be evaded by adopting a sequence of specially crafted
assumptions about the behavior of strangelets, which we consider
physically unmotivated.  It is important to remember, however, that
evading the bounds does not make dangerous strangelet production more
likely.

\subsection{A Primer on Strangelets and Strange Matter}

Strange matter is the name given to quark matter at zero temperature
in equilibrium with the weak interactions.  At and below ordinary
nuclear densities, and at low temperatures, quarks are confined to the
interiors of the hadrons they compose.

It is thought that any collection of nucleons or nuclei brought to
high enough temperature or pressure\footnote{For theoretical purposes
a better variable is chemical potential, instead of pressure.  But
either can be used.}, will make a transition to a state where the
quarks are no longer confined into individual hadrons.  At high
temperature the material is thought to become what is called a
quark-gluon plasma.  The defining property of this state is that it
can be accurately described as a gas of nearly freely moving quarks
and gluons.  One main goal of RHIC is to provide experimental evidence
for the existence of this state, and to study its properties.  At high
pressure and low temperature the material is expected to exhibit quite
different physical properties.  In this regime, it is called quark
matter.  Quarks obey the Pauli exclusion principle --- no two quarks
can occupy the same state.  As quark matter is compressed, the
exclusion principle forces quarks into higher and higher energy
states.

Given enough time (see below), the weak interactions will come
into play, to reduce this energy.  Ordinary matter is made of up ($u$)
and down ($d$) quarks, which are the lightest species (or ``flavors'')
of quarks.  The strange quark ($s$) is somewhat heavier.  Under
ordinary conditions when an $s$ quark is created, it decays into $u$
and $d$ quarks by means of the weak interactions.  In quark matter the
opposite can occur.  $u$ and $d$ quarks, forced to occupy very
energetic states, will convert into $s$ quarks.  Examples of weak
interaction processes that can accomplish this are strangeness
changing weak scattering, $u+d\to s+u$, and weak semi-leptonic decay,
$u\to s+e+\bar \nu_{e}$.  These reactions occur rapidly on a natural
time scale $\sim 10^{-14}$ sec.  When the weak interactions finish
optimizing the flavor composition of quark matter, there will be a
finite density of strange quarks --- hence the name ``strange
matter''.

The most likely location for the formation of strange matter
is deep within neutron stars, where the mammoth pressures generated by
the overlayers of neutrons may be sufficient to drive the core into a
quark matter state.  When first formed, the quark matter at the core
of a neutron star would be non-strange, since it was formed from
neutrons.  Once formed, however, the quark matter core would rapidly
equilibrate into strange matter, if such matter has lower free energy
at high external pressure.

Initially, the non-strange quark matter core and the overlaying layer
of neutrons were in equilibrium.  Since the strange matter core has
lower free energy than the overlaying neutrons, its formation disrupts
the equilibrium.  Neutrons at the interface are absorbed into the
strange matter core, which grows, eating its way outward toward the
surface.  There are two possibilities.  If strange matter has lower internal
energy than nuclear matter even at zero external pressure, the
strange matter will eat its way out essentially to the surface of the
star.  On the other hand, if below some non-zero pressure, strange
matter no longer has lower energy than nuclear matter, the conversion
will stop.  Even in the second case a significant fraction of the star
could be converted to strange matter.  The ``burning'' of a neutron
star as it converts to strange matter has been studied in
detail~\cite{Alcock:1991bw,Madsen:1991dz}.  It is not thought to
disrupt the star explosively, because the free energy difference
between strange matter and nuclear matter is small compared to the
gravitational binding energy.

In 1984, E.~Witten suggested that perhaps strange matter has
lower mass than nuclear matter even at zero external pressure
\cite{Witten}.  Remarkably, the stability of ordinary nuclei does not
rule this out.  A small lump of strange matter, a ``strangelet'',
could conceivably have lower energy than a nucleus with the same
number of quarks.  Despite the possible energy gain, the nucleus could
not readily decay into the strangelet, because it would require many
weak interactions to occur simultaneously, in order to create all the
requisite strange quarks at the same time.  Indeed, we know that
changing one quark (or a few) in a nucleus into an $s$ quark(s) ---
making a so-called hypernucleus --- will raise rather than lower the
energy.

Witten's paper sparked a great deal of interest in the physics and
astrophysics of strange quark matter.  Astrophysicists have examined
neutron stars both theoretically and observationally, looking for
signs of quark matter.  Much interest centers around the fact that a
strange matter star could be considerably smaller than a neutron star,
since it is bound principally by the strong interactions, not gravity.
A small quark star could have a shorter rotation period than a neutron
star and be seen as a sub-millisecond pulsar.  At this time there is
no evidence for such objects and no other astrophysical evidence for
stable strange matter, although astrophysicists continue to search and
speculate~\cite{Madsen:1998uh}.

Strange matter is governed by QCD. At extremely high densities
the forces between quarks become weaker (a manifestation of asymptotic
freedom) and one can perform quantitatively reliable calculations with
known techniques.  The density of strange matter at zero external
pressure is not high enough to justify the use of these techniques.
Nevertheless the success of the ordinary quark model of hadrons leads
us to anticipate that simple models which include both confinement and
perturbative QCD provide us good qualitative guidance as to the
properties of strange matter~\cite{FJ}.

Such rough calculations cannot answer the delicate question of
whether or not strange matter is bound at zero external pressure
reliably.  Stability seems unlikely, but not impossible.

Some important qualitative aspects of strange matter dynamics that figure
in the subsequent analysis are as follows:

\paragraph{Binding Systematics}\cite{FJ}

  The overall energy scale of strange matter is determined by the
   confinement scale in QCD which can be parameterized by the ``bag
   constant''.  Gluon exchange interactions between quarks provide
   important corrections.  Calculations indicate that gluon
   interactions in quark matter are, on average, repulsive, and tend
   to destabilize it.  To obtain stable strange matter it is necessary
   to reduce the value of the bag constant below traditionally favored
   values~\cite{Madsen:1998uh,FJ}. This is the reason we describe
   stability at zero external pressure as ``unlikely''.

\paragraph{Charge and flavor composition}\cite{FJ}

   If strange matter contained equal numbers of $u$, $d$ and
   $s$ quarks it would be electrically neutral.  Since $s$ quarks are
   heavier than $u$ and $d$ quarks, Fermi gas kinematics
   (ignoring interactions) would dictate that they are suppressed,
   giving strange matter a positive charge per unit baryon number,
   $Z/A >0$.

   If this kinematic suppression were the only consequence of
   the strange quark mass, strange matter and strangelets would
   certainly have positive electric charge.  In a bulk sample of
   quark matter this positive quark charge would be shielded by a
   Fermi gas of electrons electrostatically bound to the strange
   matter, as we discuss further below.  Energy due to the exchange
   of gluons complicates matters.  As previously mentioned,
   perturbation theory suggests this energy is repulsive, and tends to
   unbind quark matter.  However, gluon interactions weaken as quark
   masses are increased, so the gluonic repulsion is smaller between
   $s$-$s$, $s$-$u$ or $s$-$d$ pairs than between $u$ and $d$ quarks.
   As a result, the population of $s$ quarks in strange matter is
   higher than expected on the basis of the exclusion principle alone.
   If, in a model calculation, the strength of gluon interactions is
   increased, there comes a point where strange quarks dominate.
   Then the electric charge on strange matter becomes negative.

   Increasing the strength of gluon interactions pushes the charge of
   quark matter negative.  However it also unbinds it.  Unreasonably
   low values of the bag constant are necessary to compensate for the
   large repulsive gluonic interaction energy\footnote{Some early
   studies that suggested negatively charged strange matter for broad
   ranges of parameters were based on incorrect applications of
   perturbative QCD.}.  For this reason we consider a negative charge
   on strange matter to be extremely unlikely.

\paragraph{Finite size effects}\cite{FJ,BJ,Madsen:1994vp,GJ}

   If it were stable, strange matter would provide a rich new kind of
   ``strange'' nuclear physics~\cite{FJ,BJ,DeRujula:1984ig}.  Unlike
   nuclei, strangelets would not undergo fission when their baryon
   number grows large.  Nuclear fission is driven by the mismatch
   between the exclusion principle's preference for equal numbers of
   protons and neutrons and electrostatics' preference for zero
   charge.  In strange matter there is little mismatch: $u\approx
   d\approx s$ coincides with approximately zero charge.

   On the other hand strangelets, like nuclei, become less stable at
   low baryon number.  Iron is the most stable nucleus.  Lighter
   nuclei are made less stable by surface effects.  Surface energy is
   a robust characteristic of degenerate fermion systems.  Estimates
   suggest that strange matter, too, has a significant surface energy,
   which would destabilize small
   strangelets~\cite{FJ,BJ,Madsen:1994vp}.  The surface tension which
   makes light nuclei and water droplets roughly spherical is a well
   known manifestation of positive surface energy.  The exact value of
   $A$ below which strangelets would not be stable is impossible to
   pin down precisely, but small values of $A$ (eg.  less than 10--30)
   are not favored.

   Some very small nuclei are very stable.  The classic example is
   $^{4}$He.  The reasons for helium's stability are very well
   understood.  A similar phenomenon almost certainly does
   not occur for strangelets.  The pattern of masses for
   strangelets made of 18 or fewer quarks can be estimated rather
   reliably \cite{FJ}.  Gluon interactions are, on average,
   destabilizing.  They are most attractive for six quarks, where they
   still fail to produce a stable strange hadron.  The most bound
   object is probably the $H$, composed of $uuddss$ \cite{RLJ}.  It is
   unclear whether this system is stable enough to be detected.  On
   empirical grounds, it is certainly not lighter than the non-strange
   nucleus made of six quarks --- the deuteron.  For $2<A\le 6$, QCD
   strongly suggests complete instability of any strangelets.  Larger
   strangelets, with baryon numbers up to of order 100, have been
   modelled by filling modes in a bag
   \cite{Madsen:1994vp,GJ,Greiner:1988pc}.  These admittedly crude
   studies indicate the possible existence of metastable states, but
   none are sufficiently long-lived to play a role in catastrophic
   scenarios at a heavy ion collider.  Thus, even if it were stable in
   bulk, strange matter would be unlikely to be stable in small
   aggregates.

\paragraph{Strangelet radioactivity and metastability}~\cite{BJ,GJ}

   If strange matter is stable in bulk and finite size effects
   destabilize small strangelets, then there will likely be a range of
   $A$ over which strangelets are metastable and decay by various
   radioactive processes.  The lighter a strangelet, the more unstable
   and shorter lived it would be.  Two qualitatively different kinds
   of radioactivity concern us: baryon emission and lepton or photon
   emission.

   \begin{itemize}

   \item Baryon emission

   {\BF It might be energetically favorable for a small strangelet to
   emit} baryons (neutrons, protons, or $\alpha$ particles, in
   particular), and reduce its baryon number.  Such decays are likely
   to be very rapid.  Strong baryon emission would have a typical
   strong interaction lifetime of order $10^{-23}$ sec.  $\alpha$
   decay, which can be very slow for nuclei, would be very rapid for a
   negatively charged strangelet on account of the absence of a
   Coulomb barrier.  Weak baryon emission would be important for some
   light strangelets that must adjust their strangeness in order to
   decay.  The lifetime for weak baryon emission can be approximated
   by
   \begin{equation}
           \tau^{-1}\approx \frac{Q}{4\pi}\sin^{2}\theta_{c}G_{F}^{2}\mu^{4}
           \label{eq:1}
   \end{equation}
   where $G_{F}$ is Fermi's constant ($G_{F}=10^{-5}M_{p}^{-2}$),
   $\sin\theta_{c}$ is Cabibbo's angle, $Q$ is the Q-value of the
   decay, and $\mu$ is the quark chemical potential in strange matter.
   Reasonable choices for these parameters put $\tau$ below $10^{-8}$
   sec.
   
   Baryon emission leaves a small strangelet smaller still, and less
   stable.  Strangelets unstable against baryon emission quickly
   decay away to conventional hadrons.

   \item Lepton or photon emission

   A strangelet which is stable against baryon emission would adjust
   its flavor through a variety of weak processes until it reached a
   state of minimum energy.  The underlying quark processes include
   electron or positron emission, $(d \ {\rm or}\  s)\to u e^{-}\bar\nu_{e}$, $u\to
   (d\ {\rm or}\ s) e^{+}\nu_{e}$, electron capture, $ue^{-}\to (d\ 
   {\rm or}\ s)\nu_{e}$,
   and weak radiative strangeness changing scattering, $ud\to
   su\gamma$.  These processes are much slower than baryon emission
   because they typically have three body final states, initial state
   wavefunction factors, or other suppression factors.  Rates would
   depend on details of strangelet structure which cannot be
   estimated without a detailed model.   We would expect lifetimes to
   vary as widely as the $\beta$ decay and electron capture lifetimes
   of ordinary nuclei, which range from microseconds to longer than
   the age of the universe.

   \item Systematics of stability

   The only studies of strangelet radioactivity were done in the
   context of a rather primitive model\cite{GJ}.  Even then, some
   features emerge that would have significant implications for the
   disaster scenarios which concern us.  Specifically,

   \begin{itemize}
           \item
           Even if the asymptotic value of $Z/A$ were negative,
           there probably would exist absolutely stable strangelets with
           positive charge.  Production of such a species would terminate
           the growth of a dangerous strangelet (see below).  The 
           opposite case
           (a negatively charged strangelet in a world where $Z/A$ is
           asymptotically positive) would not present a hazard.
           \item
           Calculations indicate that the lightest (meta)stable
           strangelet can occur at a value of $A\equiv A_{\rm min}$ well
           below the onset of general stability, with no further stable
           species until some $A'\gg A_{\rm min}$.  This phenomenon occurs
           in conventional nuclear physics at the upper end of the
           periodic table, where occasional (meta)stable nuclei exist in
           regimes of general instability.  In this case a dangerous
           strangelet could not grow by absorbing matter.
    \end{itemize}

        Even though these features of strangelet stability could stop the
        growth of a negatively charged strangelet produced at RHIC, we
        cannot use them to argue for the safety of RHIC because we do not
        know how to model them accurately.
    \end{itemize}
For the sake of definiteness, we will refer to any strangelet with a
lifetime long enough to be produced at RHIC, come to rest, and be captured
in matter as ``metastable''.  To summarize:  strangelets which decay
by baryon emission have lifetimes which are generally too short to be
``metastable''.  Thus any strangelets which eventually evaporate away
do so very quickly.  On the other hand, strangelets which decay by
lepton or photon emission could be quite long lived.

\subsubsection{Searches for Strange Matter}

   In addition to the astrophysical searches reviewed in
   Refs.~\cite{Madsen:1991dz,Madsen:1998uh}, experimental
   physicists have searched unsuccessfully for stable or quasi-stable
   strangelets over the past 15 years.  Searches fall in two principal
   categories: a) searches for stable strangelets in matter; b)
   attempts to produce strangelets at accelerators.

   Stable matter searches look for stable stangelets created sometime
   in the history of our Galaxy, either in cosmic ray collisions or as
   by products of neutron star interactions.  Due to its low charge to
   mass ratio, a stable light strangelet would look like an ultraheavy
   isotope of an otherwise normal element.  For example a strangelet
   with $A\approx 100$ might have $Z=7$.  Chemically, it would behave
   like an exotic isotope of nitrogen, $^{100}$N(!)  Searches for
   ultraheavy isotopes place extremely strong limits on such
   objects~\cite{Hemmick:1989ns}.  The failure of these searches
   is relevant to our considerations because it further reduces
   the likelihood that strange matter is stable in bulk at zero
   external pressure~\cite{BlJ}.

   Accelerator searches assume only that strangelets can be produced
   in accelerators and live long enough to reach detectors. 
   Experiments to search for strangelets have been carried out at the
   Brookhaven National Laboratory Alternating Gradient Accelerator
   (AGS) and at the CERN Super Proton Accelerator (SPS).  At the AGS
   the beam species and energy were gold at an energy of {\bf 11.5
   GeV/nucleon}\cite{xu}.  At the CERN SPS the beam was lead at an
   energy of 158 GeV/nucleon \cite{kling1}.  Experiments (with less
   sensitivity) were also done at CERN with sulfur beams at an energy
   of 200 {\bf GeV/nucleon}\cite{kling2}.  In all of these experiments
   the targets were made of heavy elements (lead, platinum and
   tungsten).

   All of the experiments were sensitive to strangelets of both
   positive and negative electric charge.  All of the experiments
   triggered on the low value of $Z/A$ characteristic of strangelets.
   The experiments were sensitive to values of $| Z/A | \lesssim 0.3$,
   masses from 5 GeV/c$^{2}$ to 100 GeV/c$^{2}$, and lifetimes longer
   than 50 ns ($5\times 10^{-8}$ seconds).

   None of the experiments detected strangelet signals.
   Limits were therefore set on the possible production rates of
   strangelets with the stated properties.  The limits achieved were
   approximately less than one strangelet in $10^{9}$ collisions at
   the AGS and from one strangelet per $10^{7}$ to $10^{9}$ collisions
   at CERN energies, depending on the precise properties of the
   strangelet.

   Of course the limits obtained from previous strangelet searches
   cannot be used to argue that experiments at RHIC are safe because
   the total luminosity of earlier searches would not place a decisive
   limit on the probability of negative strangelet production at
   RHIC.  However, attempts to understand possible strangelet production
   mechanisms in these experiments figure importantly in our
   consideration of dangerous strangelet production at RHIC.

\subsection{Strangelet Production in Heavy Ion Collisions}

The lack of a plausible mechanism whereby hypothetical
dangerous strangelets might be produced is one of the weakest links in
the catastrophe scenario at a heavy ion collider.  Before discussing
production mechanisms in detail, it is worthwhile to summarize some of
the very basic considerations that make dangerous strangelet
production appear difficult.

\begin{itemize}
   \item Strangelets are cold, dense systems.  Like nuclei, they are
   bound by tens of MeV (if they are bound at all).  Heavy ion
   collisions are hot.  If thermal equilibrium is attained,
   temperatures are of order one hundred MeV or more.  The second law
   of thermodynamics fights against the condensation of a system an
   order of magnitude colder than the surrounding medium.  It
   has been compared to producing an ice cube in a furnace.

   \item $q\bar q$ pairs, including $s\bar s$ pairs, are most
   prevalent in the central rapidity region in heavy ion collisions.
   Baryon chemical potential is highest in the nuclear fragmentation
   regions.  To produce a strangelet one needs both high chemical
   potential and many $s$ quarks made as $s\bar s$ pairs.  But the two
   occur in different regions.

   \item Strangelets include many strange quarks.  The more negative
   the strangelet charge, the more strange quarks.  For example, a
   strangelet with $A=20$ and $Z=4$ would include 12 $s$ quarks if the
   number of $u$ and $d$ quarks are equal (as expected).  However, a
   strangelet with $A=20$ and $Z=-1$ would have to contain 22 $s$
   quarks.  The more strange quarks, the harder it is to produce a
   strangelet.  Thus dangerous strangelets are much harder to make
   than benign ($Z>0$) strangelets.

   \item As we have previously discussed, the smaller the strangelet,
   the less likely it is to be stable or even metastable.  The last
   several items make it clear that the larger the strangelet, the
   less likely it is to be produced in a heavy ion collision.
\end{itemize}

We find that these arguments, though qualitative, are quite
convincing.  Especially, they strongly suggest that strangelet
production is even more unlikely at RHIC than at lower-energy
facilities (e.g.\ AGS and CERN) where experiments have already been
performed.

Unfortunately, the very unlikelihood of production makes it
difficult to make a reasonable model for how it might occur, or to
make a quantitative estimate.

Two mechanisms have been proposed for strangelet production in high
energy heavy ion collisions: a) coalescence and b) strangeness
distillation.  The coalescence process is well known in heavy ion
collisions and many references relate to it.  A recent study which
summarizes data at the AGS energies has been reported
\cite{Nagle:1999yb}.  The strangeness distillation process was first
proposed by Heinz et al. and Greiner et al.\cite{GH}.

The coalescence process has been carefully studied at AGS energies
\cite{Nagle:1999yb}.  The coalescence model is most easily summarized
in terms of a penalty factor for coalescing an additional unit of
baryon number and/or strangeness onto an existing clump.  By fitting
data, Ref.~\cite{Nagle:1999yb} finds a penalty factor of 0.02 per
added baryon.  The additional penalty for adding strangeness has been
estimated at 0.2, however the data of Ref.~ \cite{Nagle:1999yb}
suggests that it might be as small as 0.03.  The model {\BF was
originally intended to estimate the probability of producing nuclei
and hypernuclei from the coalescence of the appropriate number and
types of baryons.  When it is used to estimate stranglet production,
it is assumed that the transition from hadrons to quarks occurs with
unit probability.  This is certainly a gross overestimate, since
wholesale reorganization of the quark wavefunctions is necessary to
accomplish this transition.  By ignoring this factor we obtain a very
generous {\it overestimate} of the strangelet production probability.}
Given that the probability of producing a deuteron in the collision is
about unity, this suggests that the yield of a strangelet with, for
example A=20, Z=-1, and S=22 is about one strangelet per 10$^{46}$
collisions (taking the strangeness penalty factor as 0.2).  This would
lead to a probability ${\frak p} \approx 2\times 10^{-35}$ for
producing such a strangelet at RHIC. The difficulty of producing a
(meta)stable, negatively charged strangelet (if it exists) is one of
the principal reasons we believe there is no safety problem at RHIC.

In addition, the coalescence factors are expected to decrease as the
collision energy increases.  This is because the produced particles
are more energetic, and therefore less likely to be produced within
the narrow range of relative momentum required to form a coalesced
state.  If one compares the coalescence yields at the Bevalac, the
AGS, and the CERN experiments, this expectation is dramatically
confirmed.  From the point of view of coalescence, the most favorable
energy for strangelet production is below that of the AGS.

Closely related to the coalescence model is the thermal model, in
which it is assumed that particle production reflects an equilibrium
state assumed to exist until the fireball cools and collisions cease. 
In this model the ``free'' parameters are the temperature and the
baryon chemical potential at freeze-out \cite{braun}.  Applying this
model to the AGS experimental situation gives a reasonably good
account of particle ratios, and indicates a freeze-out temperature of
140 MeV and a baryon chemical potential of 540 MeV. With these
parameters the model can predict the production probability of
strangelets with any given baryon number, charge, and strangeness. 
Braun-Munzinger and Stachel \cite{braun2} have carried out detailed
calculations for the AGS case and find very small production.  For
example, the yield of a strangelet with A=20, Z=2, and S=16 is $\sim 2
\times 10^{-27}$ per central collision.  {\BF Since central collisions
are about 0.2 of all collisions this translates into a yield of one
strangelet (with these parameters) in $2 \times 10^{27}$ collisions if
such a strangelet were stable and if we scale without change from AGS
to RHIC energy.  The yield of a negatively charged strangelet would be
much smaller still.}

As the collision energy increases, this model predicts higher
temperatures and smaller baryon chemical potentials.  The result is
that in this model  strangelet production is predicted to
decrease quickly with total center of mass energy in this model.  The
thermal model clearly favors an energy even lower than the AGS for the
optimum for producing strangelets, should they exist.

The strangeness distillation mechanism is considerably more
speculative.  It assumes that a quark gluon plasma (QGP) is produced
in the collision and that the QGP is baryon rich.  It further assumes
that the dominant cooling mechanism for the QGP is evaporation from
its surface.  Since it is baryon rich, there is a greater chance for
an $\bar{s}$ quark to find a $u$ or $d$ quark to form a kaon with
positive strangeness than for an $s$ quark to find a $\bar{u}$ or
$\bar{d}$ quark to form a kaon with negative strangeness.  The
QGP thus cools to a system containing excess $s$ quarks, which
ultimately becomes a strangelet.

This mechanism requires a collision energy sufficient to form a QGP.
RHIC should be high enough.  Many heavy ion physicists believe that
even the fixed target CERN experiments have reached a sufficient
energy and are in fact forming a QGP. If this is the case, the failure
of the CERN experiments to find strangelets argues against either the
existence of this mechanism or the existence of strangelets.  A
substantial body of evidence supports the view that a QGP is formed at
CERN energies, but a truly definitive conclusion is not possible at
present.  {\BF In any case, fits to data from the AGS and CERN, and
theoretical models suggest that the baryon density at central
rapidity, where a QGP can be formed, will decrease at RHIC.} Moreover,
there is considerable evidence that the systems formed in CERN heavy
ion collisions do not cool by slow evaporation from the surface but
rather by rapid, approximately adiabatic expansion, as is also
expected theoretically.  Altogether, the strangeness distillation
mechanism seems very unlikely to be effective for producing
strangelets at RHIC.

In summary, extrapolation from particle production mechanisms that
describe existing heavy ion collision data suggests that strangelets
with baryon number large enough to be stable cannot be produced.  With
one exception, all production models we know of predict that
strangelet production peaks at low energies, much lower than RHIC and
perhaps even lower than the AGS. The one exception is the hypothetical
strangeness distillation mechanism.  However, available data {\BF and
good physics arguments suggest} that this mechanism does not apply to
actual heavy ion collisions.

\subsection{Catastrophe at RHIC?}

What is the scenario in which strangelet production at RHIC leads to
catastrophe?  The culprit would be a stable (or long-lived,
metastable) negatively charged strangelet produced at RHIC. It would
have to be a light representative of a generic form of strange matter
with \emph{negative} electric charge in bulk.  It would have to live
long enough to slow down and come to rest in matter.  Note that the
term ``metastable'' is used rather loosely in the strangelet
literature.  Sometimes it is used to refer to strangelets that live a
few orders of magnitude longer than strong interaction time scales.
As mentioned above, we use ``metastable'' to refer to a lifetime long
enough to traverse the detector, slow down and stop in the shielding.
Since strangelets produced at high rapidity are likely to be destroyed
by subsequent collisions, we assume a production velocity below
$v_{\rm crit} = 0.1 c$\cite{DDH}. Hence it requires a lifetime
greater than $\sim10^{-7}$ sec in order to satisfy our
definition of ``metastable''.

Once brought to rest, a negative metastable strangelet would be
captured quickly by an ordinary nucleus in the environment.  Cascading
quickly down into the lowest Bohr orbit, it would react with the
nucleus, and could absorb several nucleons to form a larger strangelet.  The
reaction would be exothermic.  After this reaction its electric charge
would be positive.  However, if the energetically preferred charge
were negative, the strangelet would likely capture electrons until it
once again had negative charge.  At this point the nuclear capture and
reaction would repeat.  Since there is no upper limit to the baryon
number of a strangelet, the process of nuclear capture and weak
electron capture would continue.

There are several ways that this growth might terminate without
catastrophic consequences: First, as mentioned earlier, a stable
positively charged species might be formed at some point in the growth
process.  This object would be shielded by electrons and would not
absorb any more matter.  Second (also mentioned before), the lightest
metastable strangelet might be isolated from other stable strangelets
by many units in baryon number.\footnote{A similar barrier (the
absence of a stable nucleus with $A=8$) prevents two $\alpha$
particles from fusing in stellar interiors.}  Third, the energy
released in the capture process might fragment the strangelet into
smaller, unstable objects.  Unfortunately, we do not know enough about
QCD either to confirm or exclude these possibilities.

A strangelet growing by absorbing ordinary matter would have an
electric charge very close to zero.  If its electric charge were
negative, it would quickly absorb (positively charged) ordinary matter
until the electric charge became positive.  At that point absorption
would cease until electron capture again made the quark charge
negative.  As soon as the quark charge became negative the strangelet
would absorb a nucleus.  Thus the growing strangelet's electric charge
would fluctuate about zero as it alternately absorbed nuclei and
captured electrons.  Even though the typical time for a single quark
to capture an electron might be quite long, the number of
participating quarks grows linearly with $A$, so the baryon number of
the strangelet would grow exponentially with time, at least until the
energy released in the process began to vaporize surrounding material
and drive it away from the growing strangelet.  This process would
continue until all available material had been converted to strange
matter.  We know of no absolute barrier to the rapid growth of a
dangerous strangelet{\BF, were such an object hypothetically to 
exist and be produced.  This is why we have considered these 
hypotheses in detail to assure ourselves beyond any reasonable doubt 
that they are not genuine possibilities.}

We should emphasize that production of a strangelet with positive
charge would pose no hazard whatsoever.  It would immediately capture
electrons forming an exotic ``strangelet-atom'' whose chemical
properties would be determined by the number of electrons.  The
strange ``nucleus'' at its core would be shielded from further nuclear
interactions in exactly the same way that ordinary nuclei are shielded
from exothermic nuclear fusion.  We see no reason to expect enhanced
fusion processes involving atoms with strangelets at their core.  It
has been suggested that an atom with a strangelet at its core would
undergo fusion reactions with light elements in the environment and,
like a negatively charged strangelet, grow without limit~\cite{Wagner}.
This will not occur.  First, the strength and range of the strong
interactions between a strangelet-atom and an ordinary atom are
determined by well-known, long-range properties of the nuclear force
which are exactly the same for strangelets as for nuclei.  Second,
fusion is suppressed by a barrier penetration factor proportional to
the product of the charge on the strangelet times the charge on the
nucleus, $f\propto e^{-Z_{1}Z_{2}K}$.  The most favorable case would
be a strangelet of charge one fusing with hydrogen.  Hydrogen-hydrogen
fusion at room temperature is so rare that it is a subject of intense
debate whether it has ever been observed.  Even if
strangelet-atom-hydrogen fusion were enhanced by some unknown and
unexpected mechanism, the suppression factor that appears in the
exponent would be doubled as soon as the strangelet had acquired a
second unit of charge.  As the strangelet's charge grows each
successive fusion would be breathtakingly more suppressed.

To provide a concrete example, we have calculated the rate of fusion
of a thermalized (room temperature) strangelet with baryon number 396
(the baryon number present in the entire Au-Au collision) and
$Z=6$, with hydrogen.  Using standard and well-tested nuclear
reaction theory, we find a fusion rate of $\sim 10^{-2\times 10^{5}}$
sec$^{-1}$.

On theoretical grounds alone, as discussed above, we believe creation
of a dangerous strangelet at RHIC can be firmly excluded.  We now turn
to the important empirical evidence from cosmic rays.

\subsection{Cosmic Ray Data Relevant to the Strangelet Scenario}

It is clear that cosmic rays have been carrying out RHIC-like
``experiments'' throughout the Universe since time out of mind.  Here
we choose some specific conditions and summarize briefly the arguments
that place restrictions on dangerous strangelet production at RHIC.
We have made estimates based on cosmic ray collisions with the Moon.
We also review the astrophysical estimates in a recent paper by Dar, De
Rujula and Heinz\nobreak~\cite{DDH}.

In order to extract bounds from cosmic ray data, it is
necessary to model the rapidity distribution of strangelets.  It will
turn out that the most important distinguishing features of a
production mechanism are how it behaves at central and extreme values
of the rapidity.  Inclusive hadronic processes generally fall like a
power of the rapidity near the limits of phase space.  In light of
this, we see no reason for strangelet production to be exponentially
suppressed at $Y_{\rm min}$ and $Y_{\rm max}$.  On the other hand,
long-standing theoretical ideas and phenomenology suggest the
emergence of a ``central plateau'' away from the kinematic limits of
rapidity, along which physics is independent of the rapidity.  Insofar
as these ideas are correct, a singularity at central rapidity would
violate the principle of relativity.

So for our first model we assume a power law dependence at the
kinematic limits of rapidity, and an exponential fall off away from
the target fragmentation region, where the baryon chemical potential
decreases. By convention we take $y=0$ to be the kinematic limit
and we model the strangelet production near $y=0$ by 
\be
   \left.\frac{d\Pi}{dy}\right|_{\rm BG} = Npy^{a}e^{-by},
   \la{C.1}
\ee
where $a$ and $b$ are parameters, $N$ is a normalization constant
chosen so that $p$ is half the total strangelet production
probability per collision (the other half comes near the other
rapidity limit).  The subscript ``BG'' stands for ``best guess''.

The authors of Ref.~\cite{DDH} have made an extreme model of
strangelet production, where production is completely confined to
central rapidity.  We know of no physical motivation for this
assumption.  On the contrary, what we know about particle production
in heavy ion collisions argues against such a model. Their model can
be approximated by a $\delta$ function at central rapidity,
\be
   \left.\frac{d\Pi}{dy}\right|_{\rm DDH} = p\delta(y-Y/2),
   \la{C.2}
\ee
where $Y$ is the total rapidity interval.  Although we find such a
model impossible to justify on any theoretical grounds, we will
use this rapidity distribution when we review the work of
Ref.~\cite{DDH}.

The limits from cosmic ray considerations depend on the
assumed rapidity distribution of strangelet production, in the
following respect.  If strangelets are produced in the nuclear
fragmentation regions, then cosmic ray collisions with stationary
nuclei on the surface of the moon provide more than adequate limits on
dangerous strangelet production at RHIC. On the other hand, if
strangelets were produced only at zero rapidity in the center of mass,
then strangelets produced on the Moon would not survive the stopping
process.  Under this hypothetical --- and we believe, quite unrealistic
--- assumption the persistence of the Moon provides no useful limit on
strangelet production.

 Dar, De Rujula, and Heinz introduce a parameter, $p$, as a
simple way to compare limits obtained in different
processes\cite{DDH}.  $p$ measures the probability to make a
strangelet in a single collision with speed low enough to survive the
stopping process at RHIC. $p$ is related to the parameter $\frak{p}$
which we introduced earlier by ${\frak p}=2\times 10^{11} p$.  We will
analyse cosmic ray data in terms of $p$ and relate the results to
$\frak{p}$ when necessary.  We assume that
${p}$ is independent of the atomic mass of the colliding ions, at
least for iron and gold.  We also assume ${p}$ is the same for
RHIC and AGS energies.  A single choice of ${p}$ simplifies our
presentation.  We will discuss the qualitative differences between AGS
and RHIC energies and between collisions of different nuclear species
where they arise.  Of course our aim is to bound $\frak{p}$ far below
unity.

We begin with our neighbor, the Moon, because we know the environment
well and know the Moon is not made of strange matter.
\footnote{Collisions of cosmic rays with the outer envelopes of stars,
gaseous planets, or even terrestrial planets with atmospheres like the
earth and venus, lead overwhelmingly to collisions with light nuclei
like hydrogen, helium, etc.  This is not a likely way to make strange
matter.} The Moon has a rocky surface rich in iron.  Using the data
from Section \ref{cosmics} it is easy to calculate the rate of collisions
between
specific heavy ions on the lunar surface.

Consider a cosmic ray nucleus $A$ colliding with a nucleus $A'$ with
fractional abundance $f_{A'}$ in the lunar soil.  The total number of
collisions at energies greater than $E$ over the 5 billion year
lifetime of the moon (from \eq{B.5}) is\footnote{\Eq{C.4} was obtained
by multiplying \eq{B.5} by $\sim 15\times 10^{16}$, the number of
seconds in five billion years, and by the fractional abundance,
$f_{A'}$.  In addition, the collision cross section varies with $A$
and $A'$ like $(A^{1/3}+A'^{1/3})^{2}$.  Since the dominant
constituents of the moon are lighter than iron, the probability of a
cosmic ray interacting with iron (or gold) is higher than measured by
its fractional abundance alone.  We ignore the $A$ dependence
of the cross section because it is small, it increases the strength
of our bounds, and it complicates our equations.}
\be
    \left.N(A,E)\right|_{\rm moon} \approx 8\times 10^{29}f_{A'}
    \frac{\Gamma(A, 10\ {\rm GeV})}
    {\Gamma({\rm Fe}, 10\ {\rm GeV})}
    \left(\frac{10\ {\rm GeV }}{E}\right)^{1.7}
    \la{C.4}
\ee
Using iron, $f_{\rm Fe} = 0.012$~\cite{moon}, and the cosmic ray
abundance of iron and ``gold'', we can calculate the number of
dangerous strangelets which would have been created on the surface
of the moon in several cases of interest as a function of $p$.
\begin{itemize}
   \item [I.] {\it Dangerous strangelet production in lunar
   iron-iron collisions at AGS energies.}

   Taking $E=10$ {\bf GeV/nucleon} and $f_{\rm Fe}=0.012$ we obtain $N_{\rm
   moon}(\hbox{ Fe-Fe, AGS})\approx 10^{28}p$ for the number of
   dangerous strangelets produced on the surface of the moon in terms
   of the probability to produce one in a single collision at RHIC
   ($p$).

   \item [II.] {\it Dangerous strangelet production in lunar
   iron-iron collisions at RHIC energies.}

        Scaling $E$ to 20 {\bf TeV/nucleon}, we find $N_{\rm moon}(\hbox{ Fe-Fe,
        RHIC})\approx 2\times 10^{22}p$

        \item [III.] {\it Dangerous strangelet production in
        lunar ``gold''-iron collisions at AGS energies.}

        The penalty of demanding ``gold'' is a factor of $10^{-5}$ in
        cosmic ray flux, so $N_{\rm moon}(\hbox{ Au-Fe, AGS})\approx
        10^{23}p$.

        \item [IV.] {\it Dangerous strangelet production in lunar
        ``gold''-iron collisions at RHIC energies.}

        Scaling $E$ to 20 {\bf TeV/nucleon}, we find $N_{\rm moon}(\hbox{ Au-Fe,
        RHIC})\approx 2\times 10^{17}p$.
\end{itemize}
The Moon does not provide useful limits for targets less
abundant than iron.

The total number of collisions on the surface of the Moon is huge
compared to the number anticipated at RHIC. However, strangelets
produced with even relatively low rapidity in the lunar rest frame do
not survive subsequent collisions with nuclei in the lunar soil.
DDH model the survival probability by assuming that strangelets
with $v_{\rm crit}<0.1c$ survive and all others are torn
apart\cite{DDH}. Here, we assume a geometric strangelet dissociation
cross section which is independent of energy, and use standard methods
to calculate a survival probability.  Our results agree with those of
DDH to within a factor of $2$ for all cases of interest. Consider a
strangelet with atomic mass $A$, charge $Z$ and rapidity $y$ in the
lunar rest frame.  Its survival probability is
\bea
    P(y,A,Z) &=& \exp [-n\sigma(A)\lambda(y,Z,A)]\nonumber\\
    &=& \exp[-4.85(1+\frac{1}{3}A^{1/3})^{2}(\cosh y
    -1)A/Z^{2}]
        \la{C.5}
\eea
Here $n$ is the density of lunar soil (assuming silicon, $n= 0.5\times
10^{23}{\rm cm}^{-3}$), $\sigma(A)$ is the geometric cross section for
the strangelet to collide with a silicon nucleus, $\sigma(A)=0.4
(1+\frac{1}{3}A^{1/3})^{2}$ barns, and $\lambda(y,Z,A)$ is the
stopping distance calculated assuming that the strangelet loses energy
only by ionization, $\lambda(y,Z,A) = 242 (\cosh y -1)A/Z^{2}{\rm
cm}$.

For a representative dangerous strangelet, e.g.\ $A=20$, $Z=-1$, the
suppression factor in \eq{C.5} is very large, $P(y,20,-1) =
\exp[-350(\cosh y -1)]$, so only strangelets with $y\approx 0$
survive.  For the rapidity distribution, \eq{C.2}, chosen by DDH, all
dangerous strangelets produced at RHIC would survive stopping, but no
strangelet would survive stopping on the moon.  The more realistic
production mechanism of \eq{C.1} yields lunar suppression factors of
$3\times 10^{-3}$, $10^{-4}$, $2\times 10^{-6}$, and $5\times 10^{-8}$
when the parameter $a$ (which controls the small $y$ behavior of
$dN/dy$) is chosen as $1,2,3$ and $4$.\footnote{These estimates apply
to $A=20$, $Z=-1$.  Larger $A$ are more suppressed, but we do not
consider production of a negatively charged strangelet with $A$ much
larger than 20 to be credible.  Larger $Z$ reduces the suppression.}
However this mechanism also reduces the probability that a strangelet
produced at RHIC will survive the stopping process.  The survival
probabilities are $8\times 10^{-3}$, $8\times 10^{-3}$, $10^{-2}$, and
$2\times 10^{-2}$, for $a=1,2,3,4$ respectively.  Thus the effective
lunar suppression factors are: an enhancement of $3$ for $a=1$, no
suppression for $a=2$, suppression by $2\times 10^{-4}$ for $a=3$, and
by $3\times 10^{-6}$ for $a=4$.  Choosing a suppression factor of
$10^{-6}$ we obtain survival probabilities of $10^{22}p$ for Case I
(iron-iron at AGS energies), $2\times 10^{16}p$ for Case II (iron-iron
at RHIC energies), $10^{17}p$ for Case III (``gold''-iron at AGS
energies), and $2\times 10^{11}p$ for Case IV (``gold''-iron at RHIC
energies).

To compare with other estimates we convert these results to bounds
on $\frak{p}$, the probability of producing a dangerous strangelet at
RHIC which survives the stopping process.  The fact that the Moon has
not been converted to strange matter over its lifetime bounds
$\frak{p}$ by ${\frak p} < 2\times 10^{-11}, 10^{-5}, 2\times
10^{-6}$, and $1$ for cases I-IV respectively.
Since we believe strangelet production to be more likely at AGS
energies than at RHIC, and believe iron to be a reasonable ``heavy
nucleus'', we take the limit from Case I very seriously.   If however,
one insists on recreating exactly the circumstances at RHIC and
insists on the worst case rapidity distribution, then lunar limits are
not applicable.

DDH explore the consequences of dangerous strangelet production in
nucleus-nucleus collisions in interstellar space.  They adopt ``worst
case'' assumptions at several points.  In particular, they demand
RHIC energies and ultra heavy nuclei (gold rather than iron), and they
assume that a dangerous strangelet is produced only at zero rapidity
in the center of mass.  Given these restrictive conditions they
compute the rate at which strangelets are produced at rest relative to
the galaxy.  Taking an energy of 100 {\bf GeV/nucleon} and an abundance relative to
iron of $10^{-5}$ in \eq{B.7},\footnote{DDH assume an $E^{-2.6}$ decay
of the cosmic ray spectrum and take $\Gamma({\rm Au})/\Gamma({\rm Fe})
\approx 3\times10^{-5}$, slightly different from our choices.} we
reproduce their result, $R(100 \hbox{GeV}, \hbox{Au}) \approx
10^{-58}$.  Multiplying by the age of the galaxy ($T_{0}=10$ billion
years) and by the probability, $p$, of dangerous strangelet
production, we find the number of dangerous strangelets produced per
cm$^{3}$ in the galaxy,
\be
    N(100\ \hbox{GeV}, \hbox{Au}) = T_{0}pR(100\ \hbox{GeV}, \hbox{Au})
      = 10^{-41}p \ \hbox{cm}^{-3}.
    \la{C.6}
\ee
DDH estimate that the material contained in a volume of
$10^{57}$cm$^{3}$ is swept up in the formation of a ``typical star'',
so that the probability of a dangerous strangelet ending up in a star
is approximately $P_{\star}\approx 10^{16}p$.  They then go on to
argue that the subsequent destruction of the star would be detectable
as a supernova-like event.  Based on $P_{\star}$ and the observed rate
of supernovas, DDH limit $p$ to be less than $10^{-19}$.  This
corresponds to a limit of $2 \times 10^{-8}$ on $\frak{p}$, the
probability of producing a dangerous strangelet during the life of
RHIC. Actually, we believe that DDH have been too conservative.  Good
physics arguments indicate that lower energy collisions are more
likely to create strangelets, and iron is nearly as good a ``heavy''
ion as gold.  If we scale down $E$ from RHIC energies (100 {\bf
GeV/nucleon}) to AGS energies (4.5 {\bf GeV/nucleon}) we gain a factor
of $4\times 10^{4}$ from the $E^{-3.4}$ dependence in \eq{B.7}.  If we
replace gold by iron we gain a factor of $10^{10}$.  So the bound on
dangerous strangelet production during the RHIC lifetime is more
nearly ${\frak p}<10^{-21}$.

Finally, we point out the implications of strangelet metastability for
these arguments.  DDH have implicitly assumed that the dangerous
strangelet produced in interstellar space lives long enough to be
swept up into a protostellar nebula.  Suppose, instead, that the
dangerous strangelet was only metastable, and that it decays away by
baryon emission with a lifetime greater than $10^{-7}$ sec but much
less than the millions of years necessary to form a star.  In this
case a dangerous strangelet produced at RHIC would have time to stop
in matter, stabilize and begin to grow.  However a strangelet formed
in interstellar space would decay harmlessly into baryons
etc.\cite{WWAK}.

We have estimated baryon emission lifetimes for strangelets.  A
lifetime of $10^{-7}$ seconds is near the upper limit of our
estimates.  Since the strangelet production cross section is likely to
fall so quickly with $A$ and $S$, the strangelet most likely to be
created at RHIC would be the least stable and would likely decay on
time scales much shorter than $10^{-7}$ seconds by strong baryon
emission.  A strangelet heavy enough to have a baryon emisison
lifetime of order $10^{-7}$ seconds would be much harder to
produce at RHIC.  Still, the astrophysical argument of DDH is
compromised by the possibility of producing a metastable strangelet
with a long enough baryon emission lifetime.  Note, however, that
instability to decays which do not change baryon number (and
therefore do not lead the strangelet to evaporate) is irrelevant.
Also, note that metastability does not compromise the lunar
arguments: a metastable strangelet produced in the lunar rest frame
would have just as much time to react as one produced at RHIC.

This discussion shows the pitfalls of pursuing the ``worst case''
approach to the analysis of empirical limits.  The rapidity
distribution necessary to wipe out lunar limits is bizarre.  The
metastability scenario necessary to wipe out the astrophysical limits
seems less unphysical, but still highly contrived.  Compelling
arguments assure us that RHIC is safe.  Nevertheless, a worst case
analysis, based on arguments which bend, if not break, the laws ot
physics, leads to a situation where there is no totally satisfactory,
totally empirical limit on the probability of producing a dangerous
strangelet at RHIC.
  
In summary, we have relied on basic physics principles to tell us
that it is extremely unlikely that negatively charged strange matter
is stable, that if it is stable in bulk, it is unlikely to be stable
in small droplets, and that even small strangelets are impossibly
difficult to produce at RHIC. In addition, empirical arguments using
the best physics guidance available, as opposed to ``worst case''
assumptions, together with data on cosmic ray fluxes, bound the
probability of dangerous strangelet production at RHIC to be
negligibly small.

\section{Acknowledgments}
We would like to thank many individuals for sharing their data and
their insights with us.  We thank B.~Price, L.~Rosenberg, S.~Swordy
and A.~Westphal for references and conversations on cosmic rays,
S.~Mathur for conversations on gravitational singularities, and
K.~Hodges and B.~Skinner for references and conversations on the
composition of the moon.  We thank A.~Kent for correspondence on risk
analysis.  We are also grateful to A.~Dar, A.~DeRujula and U.~Heinz
for sending us Ref.~\cite{DDH} and for subsequent correspondence and
conversations.

 Research  
supported in part by the Department of Energy under
cooperative agreement DE-FC02-94ER40818~(W.B. \& R.L.J.) and 
 grants
DE-FG02-92ER40704, DE-FG02-90ER40562~(J.S.), and 
 DE-FG02-90ER40542~(F.W.).

\begin{frenchspacing}

\end{frenchspacing}

\begin{references}
\bibitem{JM} This paper is a revision and adaptation of a report
commissioned by Dr.~John Marburger, Director of BNL. Dr.~Marburger
orginally charged our committee to review the issues and ``to reduce
to a single comprehensive report the arguments that address the safety
of each of the speculative `disaster scenarios'''.
\end{references}
\end{document}